# Flat lens for seismic waves

**Authors:** S. Brûlé[1]*, S. Enoch[2], S. Guenneau[2]*

**Affiliations:**

[1]Ménard, Nozay, France.

[2]Aix-Marseille Université, CNRS, Centrale Marseille, Institut Fresnel, 13013 Marseille, France.

*Correspondence to: stephane.brule@menard-mail.fr, sebastien.guenneau@fresnel.fr.

**Abstract**: Seismic risk assessment leads civil engineers to define the ground motion for the design of structures which are self-adapting to this parameter. A promising way to do this would consist of creating an artificially structured soil and to counteract the most devastating components of seismic signals. Thanks to field test with a meter-scale structured soil made of a grid of cylindrical and vertical holes in the ground and a low frequency artificial source (< 10 Hz), we show our ability to control seismic waves through a flat lens reminiscent of what Veselago and Pendry envisioned for light.

**One Sentence Summary:** Veselago-Pendry lensing is observed in a sub wavelength meter-scale structured soil (2D) during seismic field-test with artificial seismic sources.

**The concept of seismic metamaterials:** The concept of photonic crystals (*1*, *2*) and metamaterials (*3*), arose from nano-scale world and electromagnetism. Typically, it designs periodic arrangements of elements with size comparable (*1*, *2*) or much smaller than the considered wavelength (typically hundreds of nanometers for optical wavelength) that acquire effective properties of materials with unusual properties or applications such as negative optical index *(4*, *5)*, or highly anisotropic materials such as hyperbolic metamaterials *(6)* or invisibility cloaking devices *(7)*. The transition from the electromagnetic to acoustic waves was made possible, in particular, thanks to phononic crystals, which are artificial handcrafted structures *(8)*. In 2012, a first full-scale experiment was realized with a non-sub wavelength 2D grid of vertical empty cylindrical holes disturbed by a 50 Hertz source *(9)*, that showed the interest to pursue research on the interaction of structured soil with seismic waves.

**The scope of application:** Structure damages due to seismic excitation are often directly correlated to local site condition (figure 1) in the form of motion amplification and/or soil liquefaction inducing ground deformation (see supplemental material for more details). This study is concerned with seismic signal compound of short wavelengths (tens to hundreds of meters) propagating and scattering in soft sediments basin (figure 1).

**Experimental setup:** we explore the low frequency range of vibrations by means of dynamic compaction sources. A test zone constituted by a regular mesh of vertical cylindrical voids was carried out near the city of Lyon (France) in September 2012. The geology of the site is compounded of alluvial and glacial deposits (sand, clay and pebbles). Thanks to a deep borehole (1 106 m) located at less than 500 m from the site (French Bureau de Recherches Géologiques et Minières), we know that the first strong seismic reflector is deeper than 400 m (coal schists). A lesser contrast impedance is observed at 170 m depth, due to denser conglomerate.

The experimental grid is made of five discontinuous lines of self-stable boreholes 2 m in diameter (figures 1 and S1). The depth of the boreholes is 5 m and the grid spacing is 7 m. The measurement of the velocity of the pressure wave in the soil is given by a preliminary seismic test, pointing the first wave arrival at various offsets from the source. We measured a velocity around 600 m/s. The artificial source consists of the fall of a 17 tons steel pounder from a height of about 12 m to generate clear transient vibrations pulses (figure 3-B). The typical waveform of the source in time-domain looks like a second order Ricker wavelet (or "Mexican hat wavelet"). The signal is characterized by a mean frequency value at 8.15 Hz ($\lambda_{P\text{-wave}} \sim 74$ m) with a range of frequencies going from 3 to 20 Hz ($30 < \lambda_{P\text{-wave}} < 200$ m). At 5 m from the impact, the peak ground acceleration is around 0.9g (where g = 9.81 m.s$^{-2}$ is the gravity of Earth) which is significant but necessary to compensate the strong attenuation versus distance in earth materials. The void grid spacing (7 m) is lower than the smallest wavelength measured. Because of the short distance between sources and sensors, we consider that body shear waves and Rayleigh waves almost arrive simultaneously. Making the assumption of an elastic soil, with a Poisson's ratio at 0.3, it turns out that the ratio of pressure and shear wavespeeds $v_p/v_s = 1.871$. We then estimate the shear wave velocity to be around 320 m/s, what is coherent with the values known for this type of soil. Let us note that, a priori, the implementation of the void grid itself, may impact the properties of the initial ground and the time sample could contain direct and, due to local stratigraphy, reflected and refracted waves (figure 2). In this case, the signal is strongly polarized in the horizontal plane and more exactly in the x-direction i.e. perpendicularly to the side of the mesh (figure S2). This information is fundamental because it corroborates that most of the energy of the source is converted into energetic surface waves *(10)*.

To capture the ground motion's field, we set 30 three-component velocimeters ($v_x$, $v_y$, $v_z$) with a corner frequency of 4.5 Hz (−3 dB at 4.5 Hz) electronically corrected to 1 Hz. The sensors were used simultaneously with a common time base and were densely set in a quarter of the investigation area (figure S1). The pounder was consecutively dropped at five different places (sources location numbers 1, 2, 4, 6 & 7 in figure S1), and 7 to 12 times (acquisition 1 to 12) at each source location. Sensors remained fixed during the whole test and the complete field of velocity (80 m x 80 m) is obtained by means of the source symmetry ($S_1 - S_6$) and the symmetry with respect to a plane passing through the x-axis (figure S1).

**Design and results:** The design of the grid of holes was defined by numerical simulation and parametric studies, using the Kirchhoff-Love plate theory dedicated to the analysis of flexural waves (see supplemental materials and figure S1). Results of the field test are presented in figures 2 and 3. Sources $S_1$ and $S_6$ located at 30 m from the long side of the grid, simulate the case of a "far field" seismic signal. Sources $S_2$ and $S_7$, at 10 m, illustrates the "near-field" case (figure S3).

In figure 2, we present first a selection of pictures from the time history seismic test with sources $S_1$ and $S_6$ located at 30 m from the long side of the grid. The impact is pointed at 1.85 s and we have selected 6 snapshots at t = 1.90, 2.106, 2.124, 2.154, 2.316 and 2.345 s. The duration of the shock is comprised between 0.3 à 0.4 s. In seismology the Husid's diagram is often used to define a characteristic time duration $t_d$ of an earthquake. Here, this value is 0.19 s.

An important step of this survey was to identify a representative physical quantity. We decided to select the normalized square velocity at each time step $v^2 = v_x^2+v_y^2+v_z^2$. Snapshots do not illustrate directly the propagation of the wavefront itself because we also observe, thanks to sensors located between the holes, a wealth of informations in the area located between the grid and the source, inside and behind the grid.

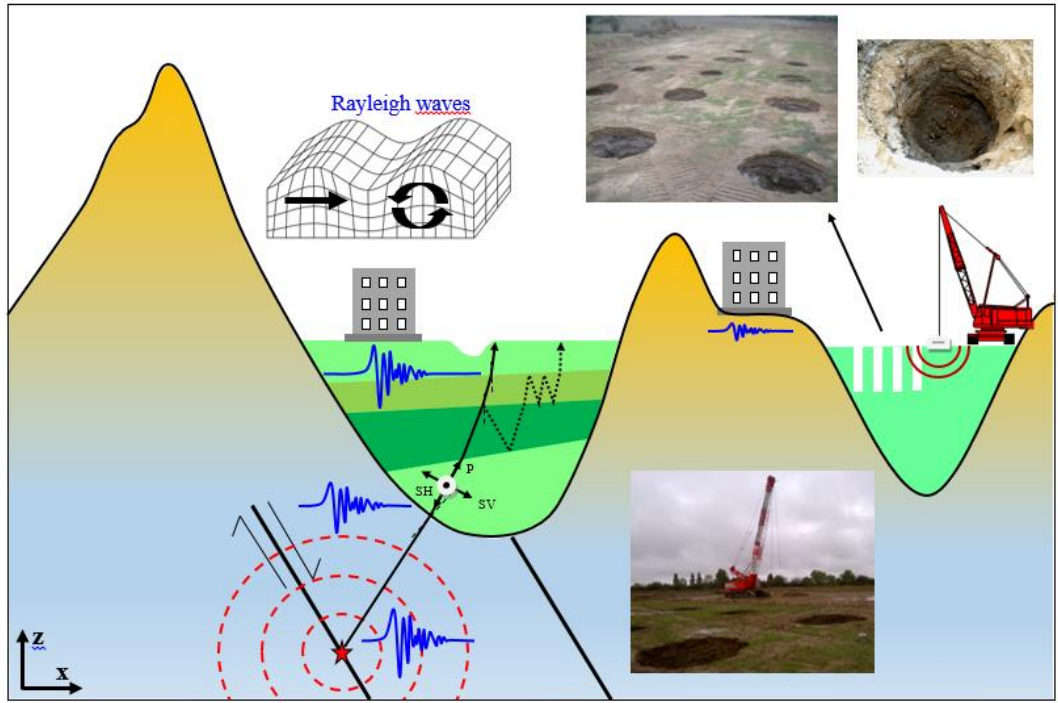

**Fig. 1.** Illustration (cross-sectional view in the x-z-plane) of the principle of seismic site effect (left) and seismic testing device (right) with pictures of the periodically structured soil made of a 7x7 m grid of 2 m diameter self-stable holes.

In time-domain, these pictures show two main phenomena. The first one is a significant wave reflection at the contact with the long-side of the grid up to 2.1 s. This phenomenon appears more markedly for the "near-field" case. See figure S4 of supplemental material. The second phenomenon, illustrated here, is the transfer and transformation of the energy inside the grid of holes. One can observe a displacement of the concentration of energy inside the grid versus time (figure 3-C) and more exactly, along the x-axis, with a succession of transient steady-states of the maxima. For the "near-field" experiment, due to the proximity of the source from the grid, the seismic signal arriving to the grid is less filtered in high frequency by soils, than in the case of the "far-field" test.

Figure 3 highligths the order of magnitude of the energy at each time step and permits to identify the origin of the signal (direct or backscattered). The total area is divided into three regions: inside (red line), behind (green line) and in front of the grid (blue line), as illustrated in figure 3 A. For each region and for the whole surface (black line), we plot the sum of the quantity $v^2$ recorded for all sensors, at a given time. This value is normalized by the maximum recorded during all the time history. One can recognize the duration of the source signal ($\Delta t = 0.375$ s) and probably from time 2.2 s, the energy of the backscattering signal. We consider that energy fields presented in figure 2 at times t = 1.90, 2.106, 2.124 and 2.154 s are representative of the direct signal coming from the source. However, fields at t = 2.316 and 2.345 s could contain a part of different reflected waves due to the geological strata. In fact, assuming a reflector at 170 m of depth, and a velocity of 600 m/s, the first time arrival of the reflected wave is 2.4 and 2.42 s respectively at the right and left long side of the grid. To identify the effect of various reflected waves at different depths, which have dominant vertical components for first time arrival, due to the small acquisition offset, we also realized the study of $v^2 = v_x^2 + v_y^2$. Results are similar.

In frequency-domain, we chose to make use of the technique inspired from "H/V" spectral ratio. The technique *(11, 12)* consists in estimating the ratio between the Fourier amplitude spectra of the horizontal to vertical components of ambient noise vibrations recorded at each sensor. This procedure is used to obtain horizontal to vertical (H/V) spectral ratios from any type of vibration signals (ambient vibrations, earthquake…). We have calculated the ratio $(|T_f(v_x)| + |T_f(v_y)|) / |T_f(v_z)|$. At a given frequency, a value greater than two, means a more important amplification of the horizontal component of the signal than the vertical one. In figure 3 D, we present the mean value of this ratio for a given area versus frequency: in front (blue curve), inside (red curve) and behind (green curve) the grid. In the range of 1 to 12 Hz, the shape of these three curves is similar with a first portion of 2 to 4 Hz of width, where the spectral ratio is greater than two. In a portion of the chart, the spectral ratio is below 2. The difference between these three curves is the location of minima and maxima. The spectral ratio is greater than 2 from 1 to 5 Hz for the inside-grid area and lower than 2 from 5 to 9 Hz. For the data recorded at the front surface, we note a frequency right-shift (1.5 to 3 Hz) of the red curve to the blue one. The range of ratio value lower than 2 is between 1 and 7 Hz for the green curve (back surface). Beforehand we verified the non-impact of the ellipticity of Rayleigh waves on the results (see supplemental material).

**Interpretation and discussion:** In time-domain, the field tests show a focusing of seismic energy inside a mesh of empty holes with sub-wavelength grid-spacing. This phenomenon may be interpreted as some form of negative refraction for surface seismic waves in a way similar to what was observed for pulse focusing of flexural waves in a 45 degree tilted square array of circular holes in a plate (*13*). Here we show Veselago-Pendry like lensing for seismic waves. In frequency-domain, we point out the capability of the grid to accentuate the filtering of the horizontal components of the signal in the range of 1 to 7 Hz, in this case. The origin of the phenomenon is the selective reflection in the front surface and probably complex changes of wave polarization at each interface. Although Rayleigh waves have more complex (elliptical) polarization than flexural waves in plates, it turns out that seismic metamaterials behave in many ways as platonic crystals which are periodically structured plates displaying some dynamic anisotropy allowing for lensing and highly directive emission effects.

**Perspectives:** This survey on a 2D-seismic metamaterial highlights the reality of the interaction of buried structures with short seismic wavelength (tens to hundreds of meters). We have shown that it is possible to achieve a significant control of seismic waves (intensity and frequency content) in artificially structured soils. Seismic metamaterial could be complementary to the existing passive earthquake engineering techniques and, with the perspective to realize such devices under and/or around a building, the anisotropy could be obtained by means of 2D (mesh of concrete or steel columns) or 3D device (cells made of walls).

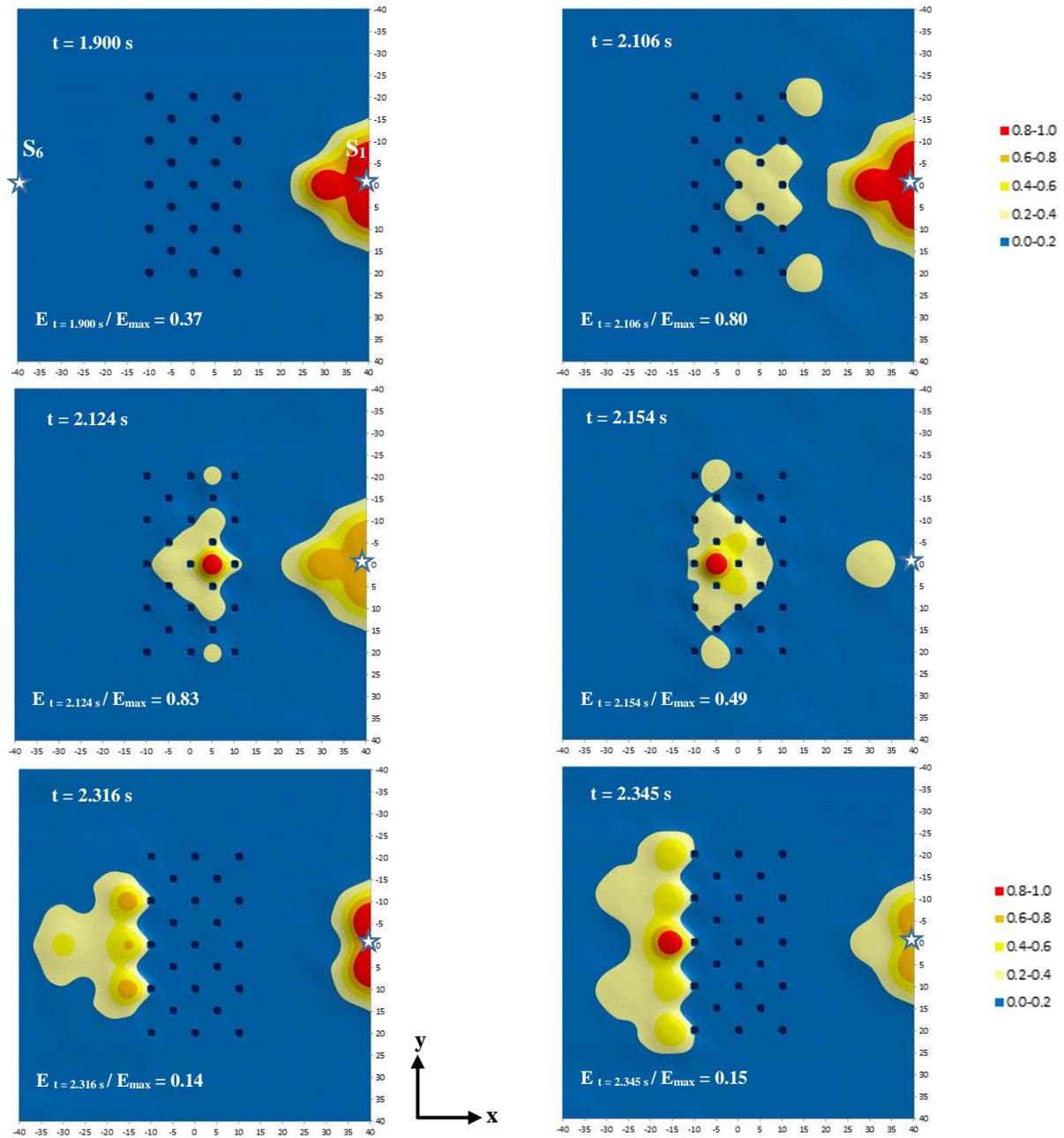

**Fig. 2.** Sources $S_1$ and $S_6$, shot#6: chronology of the x-y spatial distribution of normalized $v^2(t)$ from 1.900 to 2.345 s. Source $S_1$ (marked by a star) is located at (x=40, y=0), with coordinates in meters. Source $S_6$ is located at (x=-40, y=0). The impact is recorded at t=1.85 s at sensor F5 located at 5 m from the source.

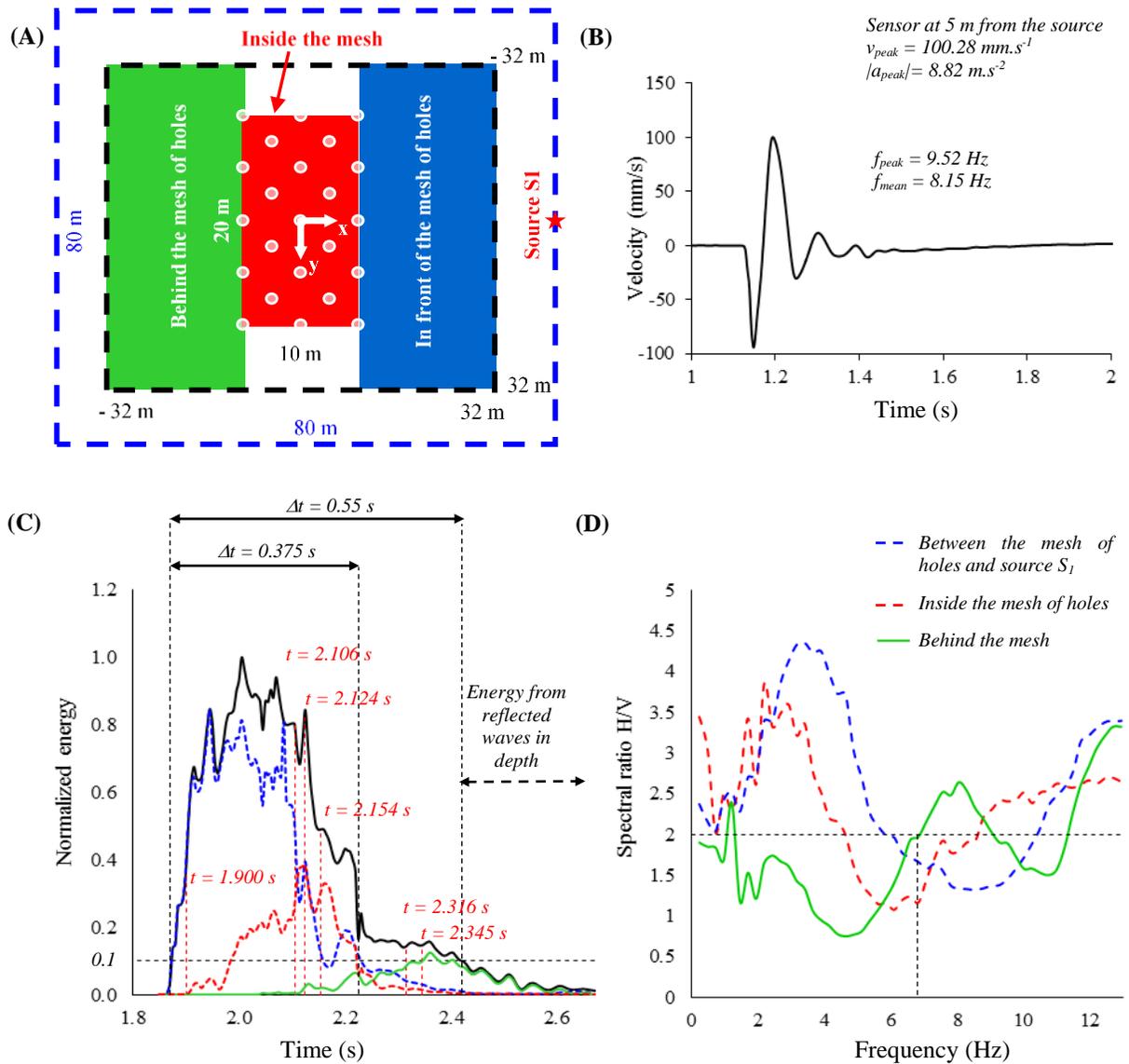

**Fig. 3.** Localization source $S_1$ and the three areas of study (A): in front, inside and behind the mesh of holes. Typical signal recorded in time domain (B) at 5 m from the source. Panel (C) shows the normalized mean energy $v^2(t)$ versus time for each surface depicted in (A). Above the black curve, we point the time of each sample illustrated in Figure 2. Panel (D) represents the spectral ratio "H/V" versus frequency for the three areas; for a given frequency, a value greater than 2, means a more important amplification of the horizontal component of the signal than the vertical one.

**References:**

1. E. Yablonovitch, Inhibited spontaneous emission in solid-state physics and electronics, *Phys. Rev. Lett.* **58**, 2059 (1987).
2. S. John, Strong localization of photons in certain disordered dielectric superlattices, *Phys. Rev. Lett.* **58**, 2486-2489 (1987).
3. D.R. Smith, J.B. Pendry, M.C.K. Wiltshire, Metamaterials and Negative Refractive Index, *Science*, **305**, 788-92 (2004).
4. V.G. Veselago, The electrodynamics of substances with simultaneously negative values of ε and μ, *Soviet Physics Uspekhi* **10** (4), 509-514 (1968).
5. J. B. Pendry, Negative refraction makes a perfect lens, *Phys. Rev. Lett.* **85**, 3966-3969 (2000).
6. A. Poddubny, I. Iorsh, P. Belov, Y. Kivshar, Hyperbolic metamaterials, *Nature Photonics*, **7** (12), 948-957 (2013).
7. J.B. Pendry, D. Schurig and D.R. Smith, Controlling Electromagnetic Fields, *Science* **312** (5781): 1789-1782 (2006).
8. R. Martínez-Sala, J. Sancho, J. V. Sánchez, V. Gómez, J. Llinares, F. Meseguer, Sound attenuation by sculpture, *Nature* **378**, 241 (1995).
9. S. Brûlé, E. H. Javelaud, S. Enoch, and S. Guenneau, Experiments on Seismic Metamaterials: Molding Surface Waves, *Phys. Rev. Lett.* **112**, 133901 (2014).
10. R.D. Woods, "Screening of surface waves in soils" (Tech. Rep. IP-804, University of Michigan, 1968).
11. M. Nogoshi, T. Igarashi, On the amplitude characteristics of microtremor (part 2), *Journal of seismological Society of Japan*, **24**, 26-40 (1971).
12. Y. Nakamura, A method for dynamic characteristics estimation of subsurface using microtremor on the ground surface, *Quaterly Report of the Railway Technical Research Institute* **30**(1), 25-30 (1989).
13. M. Dubois, E. Bossy, S. Enoch, S. Guenneau, G. Lerosey, P. Sebbah, Time-Driven Superoscillations with Negative Refraction, *Phys. Rev. Lett.* **114** (1), 013902 (2015).

**Acknowledgments:** S.G. is thankful for European funding through ERC starting ANAMORPHISM. S.B. thanks the MENARD's Earthquake Engineering Team. Supplementary Materials can be found at www.sciencemag.org All authors contributed to the concept of seismic flat lens. S.G. performed numerical simulations, S.B. conducted experiments and defined the concept of seismic metamaterials for earthquake engineering.

# Supplementary Materials for

## Flat lens for seismic waves

S. Brûlé, S. Enoch, S. Guenneau.

correspondence to: stephane.brule@menard-mail.fr , sebastien.guenneau@fresnel.fr.

**This PDF file includes:**

SupplementaryText
Figs. S1 to S4

**Supplementary Text:** We first present seismic site effect which is a major concept in seismology and earthquake engineering, in particular, to identify the range of applicability of seismic metamaterials. Then we detail our numerical modelling and results obtained (figure S1). To facilitate the interpretation of results presented in the main article, we recall the particularity of Rayleigh waves in terms of ellipticity and we present the specific analysis of the ground motion polarization, thanks to data recorded at 10 m from the impact (figure S2). An overview of the field-test is depicted in figure S3 (holes, sensors, sources). Finally, we describe the additional results obtained during the "near-field" test, with a source located at 10 m from the long side of the grid of holes (figure S4).

**Seismic site effect:** When seismic waves propagate through sediment layers (figure 1) or scatter on strong topographic irregularities, refraction/scattering phenomena may strongly increase the amplitude and the duration of the ground motion. Body waves (compressional and shear waves) propagate from the seismic source to the Earth surface and for surface waves, it is fundamental to distinguish long period surface waves (low frequency, i.e. < 1 Hz) travelling along the Earth's surface, on the crust, and short ones (< 10 Hz) mainly generated in case of site effects described above or by human activities at the Earth surface. A fundamental fact is the low value of surface wave velocity, generated by natural seismic source or construction work activities, in superficial and under-consolidated recent subgrade: less than 100 m/s to 400 m/s for shear wave. In these soil layers, considering the 0.1 to 10 Hz frequency range, wavelengths of induced surface waves are shorter than direct P and S waves ones: from a few meters to hundreds of meters. This order of wavelength is similar to those of buildings. This is the reason why we can expect building's resonance phenomena to occur with some soil in case of earthquakes (such as in the 2009 L'Aquila earthquake in Italy or more recently, in the 2015 Kathmandu earthquake in Nepal), which makes it possible to conceive seismic metamaterials whose size could be similar to those of the building project.

**Numerical results:** Let us now describe some numerical simulations which provide some phenomenological understanding of wave phenomena at stake in the experiments led by the Ménard company. Much has been said about control of light, sound, water, or shear (SH) waves (*1*, *2*, *3* and *4*) using the rich behavior encapsulated by the dispersion curves of bi-periodic structures, modeled by a Helmholtz equation, up to minor changes in the normalization of material parameters, and choice of boundary conditions (e.g. Dirichlet or Neumann for clamped or freely vibrating inclusions in the context of SH waves). However, when one moves into the area of elastic waves, governed by the Navier equations, it is no longer possible to reduce the analysis to a single scalar partial differential equation (PDE), as shear and pressure waves do couple at boundaries. There is nevertheless, the simplified framework of the Kirchhoff-Love plate theory (5) that allows for bending moments and transverse shear forces to be taken into account via a fourth-order PDE for the out-of-plane plate displacement field. This plate theory is a natural extension of the Helmholtz equation to a generic model for flexural wave propagation through any spatially varying thin elastic medium. It offers a very convenient mathematical model for any physicist wishing to grasp (some of) the physics of seismic metamaterials using earlier knowledge in photonic or platonic crystals. However, while the Helmholtz equation can, with appropriate notational and linguistic changes, hold for acoustic, electromagnetic, water or out-of-plane elastic waves and so encompasses many possible applications, the Kirchhoff-Love plate theory is dedicated to the analysis of flexural waves (it does not model propagation of in-plane elastic waves in platonic crystals). Nonetheless, if one would like to make useful analogies with photonic and phononic crystals, the biharmonic operator provides us with a playground for modeling surface elastic waves in structured soils:

$$\left(\frac{\partial^2}{\partial x_1^2} + \frac{\partial^2}{\partial x_2^2} + \Omega\right)\left(\frac{\partial^2}{\partial x_1^2} + \frac{\partial^2}{\partial x_2^2} - \Omega\right)u = 0$$

Here, $\Omega^2 = \frac{12(1-\nu^2)\rho\omega^2}{Eh^2}$, where $\rho$, h, E, $\nu$ are density, thickness, Young's modulus and Poisson's ratio of the plate, respectively, and $\omega$ is the angular wave frequency; the plate contains an array of stress-free inclusions, which is an approximate model for air-holes.

It is possible to show using high-frequency homogenization (6) that the solution u satisfies the following effective dispersion relation $\Omega \sim \Omega_0 - \left(T_{ij}/2\Omega\Omega_0\right)\kappa_i\kappa_j$ where $T_{ij}$ is a rank-2 tensor encompassing the (frequency dependent) effective elastic parameters, $\kappa_i$ is a component of the Bloch wavenumber and $\Omega_0$ is the standing wave frequency at the Brillouin zone edge 0, $-\pi/2$, $\pi/2$ depending on the location in the Brillouin zone about which the asymptotic expansion originates. For symmetry reasons, only diagonal components $T_{ii}$ are non zero. Besides from that, when $T_{11}T_{22}<0$ the effective dispersion relation describes some hyperbolic type wave propagation i.e. waves propagate like in the geometrical ray optics (7). This is exactly what happens in figure S1 A where a point source gives rise to an image inside and outside the array of air-holes. On the other hand, when $|T_{11}|<<|T_{22}|$, one observes some cross like wave propagation like in figure S1 B. This dynamic anisotropy is of foremost importance in seismic metamaterials as it enables to mould the flow of surface elastic waves.

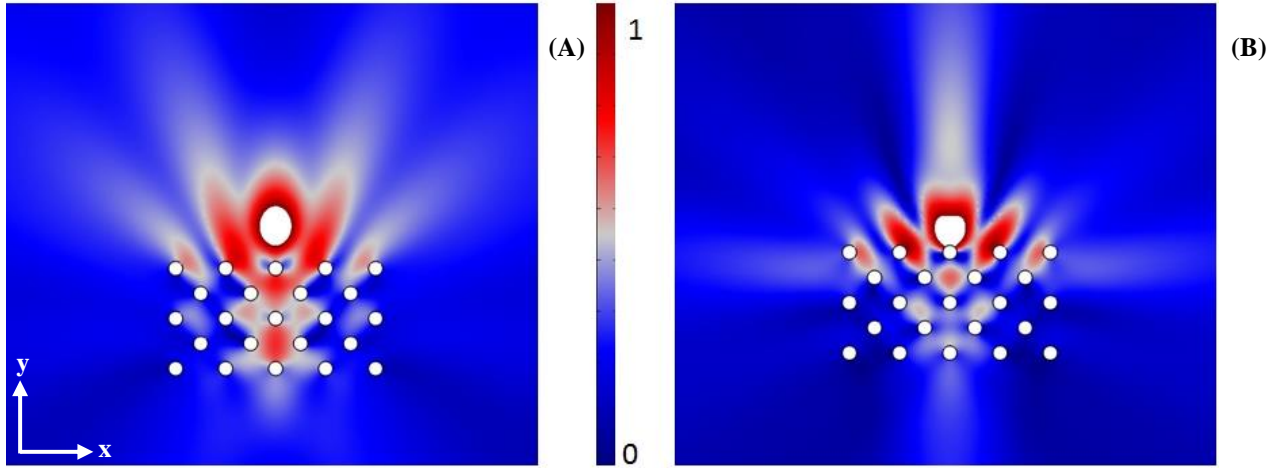

**Fig. S1**. Numerical simulations for a time-harmonic point source placed near freely vibrating inclusions in a thin plate with lensing (**A**) and guiding (**B**) effects.

**Ellipticity of Rayleigh waves:** In the left part of figure S2, we present the velocity data (x, y, z) recorded at 10 m from the impact. In the right part, we show the related plots of particle motion in horizontal (x-y) plane and vertical (x-z) plane. In this case, the signal is strongly polarized in the horizontal plane and more exactly in the x-direction i.e. perpendicularly to the side of the mesh. This information is fundamental because it corroborates that most of the energy of the source is converted into energetic surface waves.

This ellipticity may complicate the interpretation of results and the effectiveness of the grid of holes. As for velocity dispersion, the Rayleigh wave motion is frequency dependent and such ellipticity function can be represented by the Fourier transform ratio of the horizontal and the vertical components of motion versus frequency. The shape of this function could be "flat", if the velocity of compressional and shear waves for first superficial strata do not vary brutally versus depth at each interface. The preliminary seismic test realized on the virgin ground, confirm the flatness of this function in the range of 2 to 11 Hz, in accordance with the geological data.

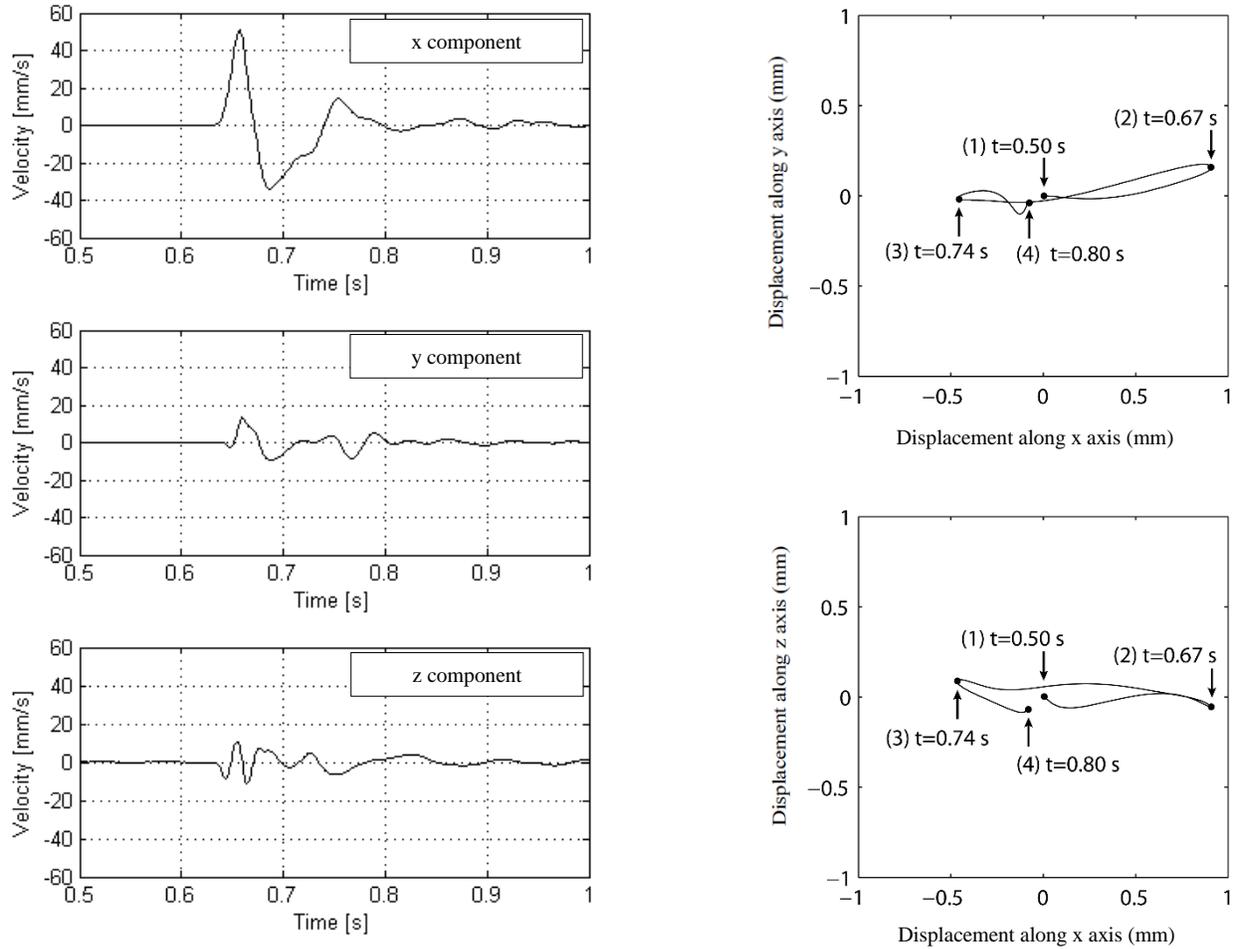

**Fig. S2**. Left, velocity versus time recorded at 10 m from the pounder impact during the preliminary seismic test (acquisition n°1). Right, displacement's orbit (top, from above, i.e. y-component versus x-component and bottom, from the side, i.e. z-component versus x).

**Geometrical characteristics of the device:** In figure S3, we present an overview of the field-device. The experimental grid is made of five discontinuous lines of self-stable boreholes 2 m in diameter. The depth of the boreholes is 5 m and the grid spacing is 7 m. The spacing between sensors is 10 m in vertical and horizontal direction, except in the grid of holes where the vertical spacing is 5 m. The source is located at 10 m ("near-field" configuration $S_1 - S_6$) or at 30 m ("far field" configuration $S_2 - S_7$) from the long side of the grid.

**"Near-field" results:** In figure S4, we present a selection of snapshots illustrating the field of $v^2$, sampled from the time history of the seismic test. Sources are $S_2$ and $S_7$, located at 10 m from the long side of the grid. The impact is pointed at 1.85 s and we have selected 4 pictures at t = 1.887, 1.909, 2.069 and 2.227 s. These pictures are selected to illustrate the strong reflection of the signal during all the duration of the impact (around 0.3 s), with two refocalization effects: one inside the grid around t = 2.069 s and a second one around t = 2.227 s. Due to the proximity of the source from the grid, the seismic signal arriving to the grid is less filtered in high frequency by soils, than in the case of the "far-field" test.

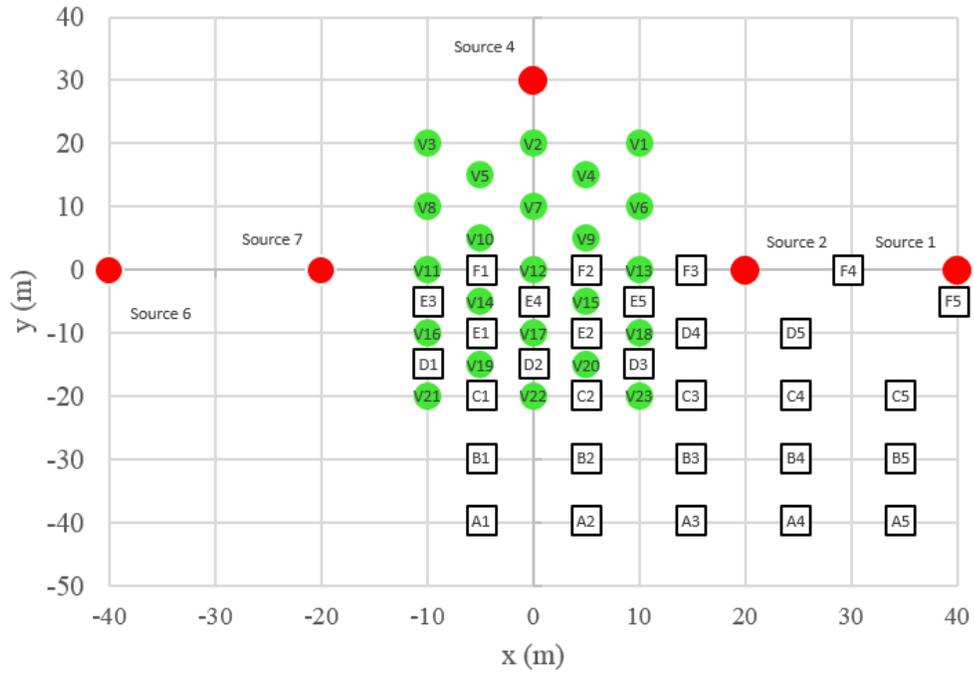

**Fig. S3.** Overview of the field test: grid of holes (green disks), mesh of sensors (black rectangles) and sources' positions (red disks).

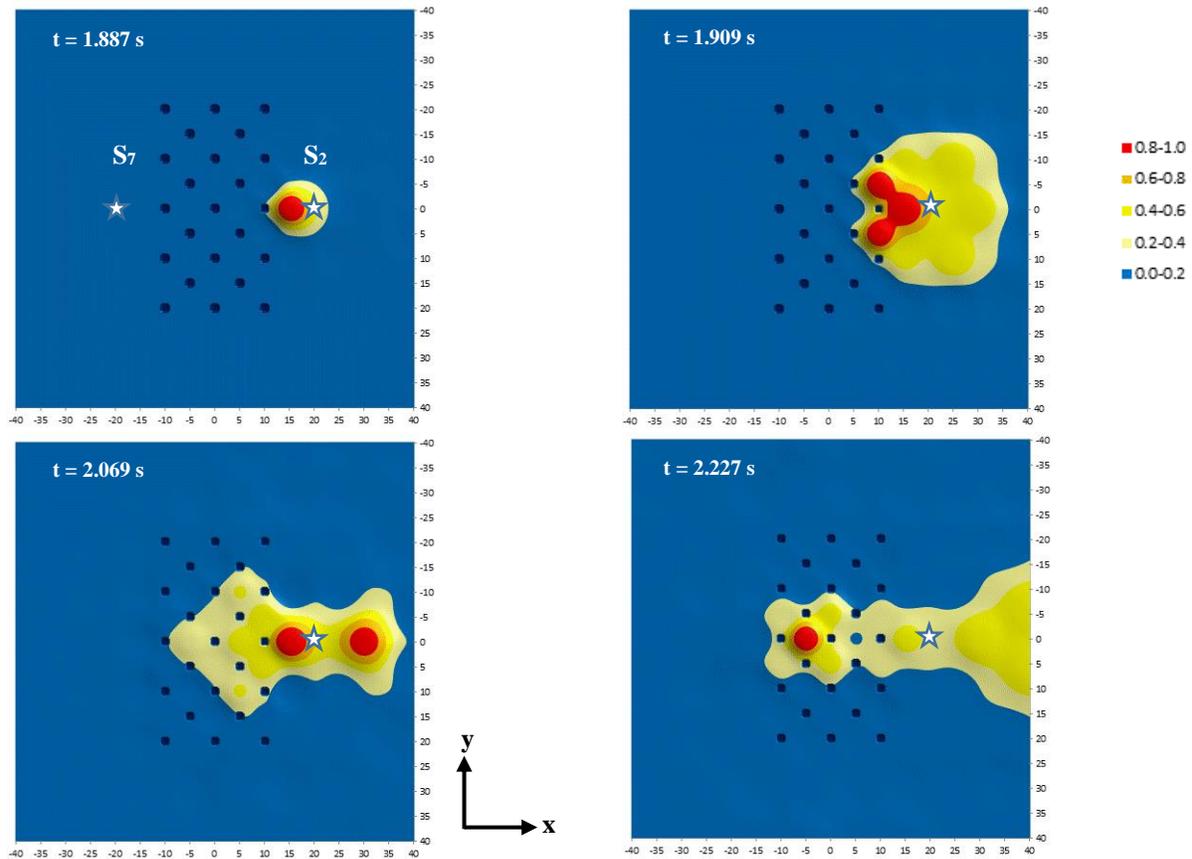

**Fig. S4**. Sources $S_2$ and $S_7$, shot#3: chronology of the x-y spatial distribution of normalized $v^2(t)$ from 1.887 to 2.227 s. Source $S_2$ is located at (x=20, y=0), with coordinates in meters. Source $S_7$ is located at (x=-20, y=0). The impact (marked by a star) is recorded at t=1.87 s at sensor F3 which is located at 5 m from the source.

**Additional references:**

1. P. Markos, C.M. Soukoulis, *Wave propagation: From electrons to photonic crystals and left-handed materials* (Princeton University Press, 2008).
2. S. Anantha Ramakrishna, T.M. Grzegorczyk, *Physics and Applications of Negative Refractive Index Materials* (CRC Press and SPIE Press, 2008).
3. P. Deymier, *Acoustic metamaterials and Photonic Crystals* (Springer Verlag, 2013).
4. R. Craster, S. Guenneau, *Acoustic metamaterials: negative refraction, imaging, lensing and cloaking* (Springer Verlag, 2013).
5. K.F. Graff, *Wave motion in elastic solids* (Dover Publications Inc., 1991).
6. T. Antonakakis, R. V. Craster, High frequency asymptotics for microstructured thin elastic plates and platonics, *Proc. R. Soc. Lond A*, 468, 1408-1427 (2012).

7. T. Antonakakis, R. V. Craster, S. Guenneau, Moulding and shielding flexural waves in elastic plates *Euro. Phys. Lett.* **105** (5), 54004 (2014).